\documentstyle[a4,amssymb]{article}

\begin{document}

\title{PUZZLE OF THE CONSTANCY OF FUNDAMENTAL CONSTANTS}
\author{D.~A. Varshalovich, A.~Y. Potekhin, and A.~V. Ivanchik \\
{\it Ioffe Physical-Technical Institute, 194021 St.\,Petersburg, Russia}
}
\renewcommand{\today}{}
\maketitle

\begin{abstract} 
We discuss experiments and observations aimed at
testing the possible space-time variability of 
fundamental physical constants, predicted by the modern theory.
Specifically, we consider two of the dimensionless physical parameters 
which are important for atomic and molecular physics:
the fine-structure constant and the electron-to-proton mass ratio.
We review the current status of such experiments and
critically analyze recent claims of a detection of the
variability of the fine-structure constant on the cosmological time scale.
We stress that such a detection remains 
to be checked by future experiments and observations.
The tightest of the firmly established
upper limits read that the considered
constants could not vary by more than 0.015\%
on the scale $\sim10^{10}$ years.



\end{abstract}

\bigskip

\section{Introduction}
Possible variability of fundamental physical constants
is one of the topical problems of contemporary physics. 
The modern theory (Supersymmetric Grand Unification Theory -- SUSY GUT, 
Superstring models, etc.)
has established that the coupling constant values
which characterize different kinds of interactions 
(i) are ``running" with the energy transfer
and
(ii) may be different in different regions of the Universe and 
vary in the course of cosmological evolution
(e.g., Ref.~\cite{Okun91}).
The energy dependence of the coupling parameters
has been reliably confirmed by high-energy experiments
(see, e.g., Ref.~\cite{Okun}),
whereas the space-time variability
of their low-energy limits so far escapes detection.

Note that a numerical value of any dimensional physical
parameter depends on arbitrary choice of physical units.
In turn, there is no way to determine the units in a remote space-time
region other than through the fundamental constants.
Therefore it is meaningless to speak of a variation of 
a dimensional physical constant without specifying which of the other 
physical parameters are
{\it defined\/} to be invariable.
Usually, while speaking of variability of a
dimensional physical parameter, one {\it implies\/}
that {\it  all\/} the other fundamental constants are fixed.
So did Milne \cite{Milne} and Dirac \cite{Dirac}
in their pioneering papers devoted to a possible change
of the gravitational constant $G$.
More recently, a number of authors 
considered cosmological theories with a time varying speed of light $c$ 
(e.g., Ref.~\cite{Albrecht}). However,
if we adopt the standard definition of meter
\cite{Petley} as the length of path traveled by 
light in vacuum in 1/299\,792\,458 s,
then $c=2.997\,924\,58\times10^{10}{\rm~cm\,s}^{-1}$ identically.
Similarly, one cannot speak of variability of the
electron mass $m_e$ or charge $e$ while using
the Hartree units ($\hbar=e=m_e=1$), most natural in atomic physics. 
Thus, only {\it dimensionless\/} combinations of the physical parameters
are truly fundamental, and only such combinations will be considered 
hereafter.

At present, the most promising candidate for the theory which 
is able to unify gravity with all other interactions is the 
Superstring theory, which treats gravity in a way consistent with 
quantum mechanics. All versions of the theory predict
existence of the dilaton -- a scalar partner to the tensorial
graviton. Since the dilaton field $\phi$ is generally not constant,
the coupling constants and masses of elementary particles, being
dependent on $\phi$, should vary in space and time. Thus, 
the existence of a weakly coupled massless dilaton entails small, 
but non-zero, observable consequences such as 
Jordan--Brans--Dicke-type deviations from General Relativity
 and cosmological variations of the gauge
coupling constants \cite{DP}. These variations depend
on cosmological evolution of the dilaton field and may be
non-monotonous as well as different in different space-time regions.

In this paper, we focus on the space-time variability
of the low-energy limits of two fundamental constants which 
are of paramount importance for atomic and molecular spectroscopy: \\
(i) the fine-structure constant $\;\alpha=e^2/\hbar c\;$ 
(Sommerfeld parameter), \\
(ii) the electron-to-proton  mass ratio $\;\mu=m_e/m_p\;$
(Born--Oppenheimer parameter).

   The next section presents a compendium 
of the basic methods allowing
   one to obtain restrictions on possible variations of fundamental
   constants. In Sects.~\ref{sect-alpha} and \ref{sect-mu}, we 
consider recent estimates of the values of $\alpha$ and $\mu$,
respectively, at cosmological redshifts $z=1$--4 
which correspond to epochs $\sim 7$--13 billion years ago. 
Conclusions are given in Sect.~\ref{sect-concl}.

\section{Tests of variability of fundamental constants}

Techniques used to investigate time
   variation of the fundamental constants may be divided into
   extragalactic and local methods. The latter ones include
   astronomical methods related to the Galaxy and the Solar system,
   geophysical methods, and laboratory measurements.

\subsection{Local tests}

\subsubsection{Laboratory measurements}
\label{sect-lab}

Laboratory tests are based on comparison of different 
frequency standards, depending on different combinations of the
fundamental constants. Were these combinations changing
differently, the frequency standards would eventually discord
with each other.
   An interest in this possibility has been repeatedly excited
since relative frequency drift was observed by
several research groups using long term comparisons of 
different frequency standards.
For instance, a comparison of frequencies of He-Ne/CH$_4$ lasers,
H masers, and Hg$^+$ clocks with a Cs standard 
\cite{Domnin,Demidov,Breakiron,Prestage} has revealed relative drifts.
Since the considered frequency standards have a different dependence 
on $\alpha$ via relativistic contributions of order $\alpha^2$,
the observed drift might be attributed to changing of the fine-structure 
constant. However, the more modern was the experiment,
the smaller was the drift.
Taking into account that the drift may be also
related to some aging processes in experimental equipment,
Prestage et al.\ \cite{Prestage} concluded that
the current laboratory data provide only an upper limit
$|\dot\alpha/\alpha|\leq3.7\times10^{-14}{\rm~yr}^{-1}$.

\subsubsection{Analysis of the Oklo phenomenon}

   The most stringent limits to variation of the fine-structure constant
$\alpha$ and the coupling constant of the strong interaction
 $\alpha_s$ have been originally inferred 
by Shlyakhter \cite{Shlyakhter}
from results of an analysis of the isotope ratio $^{149}$Sm/$^{147}$Sm 
in the ore body of the Oklo site in Gabon, West Africa. This
ratio turned out to be considerably lower than the standard one, 
which is believed to have occurred due to operation
of the natural uranium fission reactor about $2\times 10^9$ yr
ago in those ores. One of the nuclear reactions in the fission chain
was the resonance capture of neutrons by $^{149}$Sm nuclei.
Actually, the rate of the neutron capture reaction 
is sensitive to the energy of the relevant nuclear resonance level 
$E_r$, which depends on the strong and electromagnetic interaction.
Since the capture has been efficient $2\times 10^9$ yr
ago, in means that the position of the resonance has not shifted
by more than it width (very narrow) during the elapsed time.
At variable $\alpha$ and invariable $\alpha_s$ (which is just a model 
assumption), the shift of the resonance level would be determined by 
changing the difference between the Coulomb energies of 
the ground-state nucleus $^{149}$Sm 
and the nucleus $^{150}$Sm$^*$ excited to the level $E_r$. 
Unfortunately, there is no experimental data for
the Coulomb energy of the excited $^{150}$Sm$^*$ in question. 
Using order-of-magnitude estimates,
Shlyakhter \cite{Shlyakhter} concluded that 
$|\dot\alpha/\alpha|\lesssim10^{-17}{\rm~yr}^{-1}$. 
From an opposite model assumption
that $\alpha_s$ is changing whereas $\alpha=$constant, he derived a bound
$|\dot\alpha_s/\alpha_s|\lesssim10^{-19}{\rm~yr}^{-1}$.

   Damour and Dyson \cite{DD} performed a more careful analysis,
which resulted in the upper bound
$|\dot\alpha/\alpha|\lesssim7\times10^{-17}{\rm~yr}^{-1}$.
They have assumed that the Coulomb energy difference
between the nuclear states of $^{149}$Sm and $^{150}$Sm$^*$
in question is not less than that between the {\em ground\/} states 
of $^{149}$Sm and $^{150}$Sm. The latter energy difference has been 
estimated from isotope shifts and equals $\approx 1$ MeV. 
However, it looks unnatural that a weakly bound neutron ($\approx 0.1$ eV), 
captured by a $^{149}$Sm nucleus to form the highly excited
   state $^{150}$Sm$^*$, can so strongly
   affect the Coulomb energy. Moreover, heavy excited
nuclei often have Coulomb energies smaller than those for
their ground states (e.g., Ref.~\cite{Kalvius}). This indicates the
   possibility of violation of the basic assumption involved 
in Ref.~\cite{DD}, and therefore
   this method may possess a lower actual sensitivity.
Furthermore,
a correlation between $\alpha$ and $\alpha_s$
 (which is likely in the frame of modern theory)
might lead to considerable softening 
of the above-mentioned bound,
as estimated by Sisterna and Vucetich \cite{Sisterna}.

\subsubsection{Some other local tests}

Geophysical, geochemical, and paleontological data impose constraints 
on a possible changing of various combinations 
of fundamental constants over the past history 
of the Solar system, however most of these constraints are very indirect. 
A number of other methods are based on stellar and planetary
models. The radii of the planets and stars and the reaction rates
in them are influenced by values of the fundamental constants,
which offers a possibility to check variability of the constants
by studying, for example, lunar and Earth's secular accelerations.
This was done using satellite data, tidal records, and ancient
eclipses. Another possibility is offered by analyzing the data
on binary pulsars and the luminosity of faint stars. 
Most of these have relatively low sensitivity.
Their common weak point is the dependence
on a model of a fairly complex phenomenon, involving many physical effects.

An analysis of natural long-lived
$\alpha$- and $\beta$-decayers in geological minerals and meteorites
is much more sensitive. For instance, a strong bound, 
$|\dot\alpha/\alpha| < 5\times10^{-15}{\rm~yr}^{-1}$,
was obtained by Dyson \cite{Dyson} 
from an isotopic analysis of natural $\alpha$- and $\beta$-decay products 
in Earth's ores and meteorites.

Having critically reviewed the wealth of the local tests,
taking into account possible correlated synchronous changes of different
physical constants, Sisterna and Vucetich \cite{Sisterna} 
derived restrictions on possible variation rates of individual physical 
constants for ages $t$ less than a few billion years ago, 
which correspond to cosmological redshifts $z \lesssim 0.2$. 
In particular, they have arrived at the estimate
$\dot{\alpha}/\alpha = (-1.3 \pm 6.5)\times 10^{-16}\ {\rm yr}^{-1}$.

All the local methods listed above give estimates for only
a narrow space-time region around the Solar system.
For example, the epoch of the Oklo reactor ($1.8\times10^9$ years ago)
corresponds to the cosmological redshift $z \approx 0.1$.
These tests cannot be extended to earlier evolutionary stages of 
the Universe, because the possible variation of 
the fundamental constants is, in general,
unknown and may be oscillating \cite{Marciano,DP}. 
Another investigation is needed for higher cosmological redshifts.

\subsection{Extragalactic tests}

Extragalactic tests, in contrast to the local ones,  concern 
values of the fundamental constants in distant areas of the early 
Universe. 
A test which relates to the earliest epoch is based on
the standard model of the primordial
nucleosynthesis. The amount of $^4$He produced in the Big Bang is
mainly determined by the neutron-to-proton number ratio at the
freezing-out of n$\leftrightarrow$p reactions. The freezing-out
temperature $T_f$ is determined by the competition between the
expansion rate of the Universe and the $\beta$-decay rate. A
comparison of the observed primordial helium mass fraction,
$Y_p=0.24\pm0.01$, with a theoretical value allows one to obtain
restrictions on the difference between the neutron and proton
masses at the epoch of the nucleosynthesis and, through it,
to estimate relative variation of the curvature radius $R$ of
extra dimensions in multidimensional Kaluza--Klein-like theories
which in turn is related to the $\alpha$ value \cite{Kolb,Barrow87}.
However, as noted above, different coupling constants might
change simultaneously. For example, increasing 
the constant of the weak interactions $G_F$ would cause
a weak freezing-out at a lower temperature, hence a decrease in
the primordial $^4$He abundance. This process would compete with
the one described above, therefore, it reduces sensitivity of the
estimates. Finally, the restrictions would be
different for different cosmological models since the expansion
rate of the Universe depends on the cosmological constant $\Lambda$.

The most unambiguous estimation of the atomic and molecular constants
at early epochs and in distant regions of the Universe
can be performed using the extragalactic spectroscopy. 
Accurate measurements of the wavelengths in spectra of distant objects
 provide 
quantitative constraints on the variation rates of the physical constants.  
This opportunity has been first noted and used by
Savedoff \cite{Savedoff}, and in recent years exploited by
many researchers
(see, e.g., Refs.~\cite{SpSciRev,IPV} and references therein).
At present, the extragalactic spectroscopy enables one to 
probe the physical conditions in the Universe 
up to cosmological redshifts $z\lesssim4$,
which correspond, by order of magnitude, to the scales 
$\sim10^{10}$ yr in time and $\sim 10^9$ parsec in space.
In the following sections, we review briefly the studies of the space-time
variability of the fine-structure constant $\alpha$ 
and the electron-to-proton mass ratio $\mu$,
based on the latter method.

\section{Non-variability of $\alpha$}
\label{sect-alpha}

We have already mentioned in Sect.~\ref{sect-lab}
that several laboratory tests
hinted at a tentative time variation of $\alpha$,
but were later refuted 
by measurements at a higher level of accuracy.
A similar situation has occurred for 
extragalactic tests at larger space-time scales.

Bahcall and Schmidt \cite{Bahcall} were the first to use 
spectral observations of distant quasars to set 
a bound on the variability of the fine-structure constant.
They have obtained an estimate 
$\Delta\alpha /\alpha = (-2 \pm 5)\times 10^{-2}$ at $z = 1.95$.
Later statistical analyses \cite{Levshakov92,Levshakov93}
of fine-structure doublet lines 
in quasar spectra
appeared to indicate a tentative variation of $\alpha$ 
(of the order of $\sim0.3\%$ at the cosmological redshift $z\sim2$).
However, this tentative variation 
has been shown to result from a statistical bias
\cite{PV94}. 

Another statistical examination of the fine-doublet wavelengths
of absorption lines 
in quasar spectra \cite{SpSciRev} indicated a tentative 
(at the 2--3$\sigma$ level) variability
of $\alpha$ values by $\sim0.1\%$ over the celestial sphere 
(as function of angle) at redshifts $z\sim2$--3.
However, this result has not been confirmed by 
a later analysis \cite{IPV}, 
which was based on higher-quality spectra and 
yielded an order of magnitude higher precision.

Quite recently, Webb et al.\ \cite{Webb} have 
estimated $\alpha$ by comparing wavelengths of Fe{\sc\,ii}
and Mg{\sc\,ii} fine-splitted spectral lines
in extragalactic spectra and in the laboratory.
Their result suggests a time-variation of $\alpha$ 
at the incredibly high accuracy level of $\sim 10^{-3}\%$:
the authors' estimate reads
$\Delta\alpha/\alpha=(-1.9\pm0.5)\times10^{-5}$ at $z=1.0$--1.6.
Note, however, 
two important sources of a possible systematic error
which could mimic the effect:
(a) Fe{\sc\,ii} and Mg{\sc\,ii} lines used
are situated in different orders of the echelle-spectra,
so relative shifts in calibration of the different orders
can affect the result of comparison,
and (b) if the isotopic composition varies 
during the evolution of the Universe, then the average
doublet separations should vary due to the isotopic shifts. 
Were the relative abundances of Mg isotopes
changing during the cosmological evolution,
the Mg{\sc\,ii} lines would be subject to an additional
$z$-dependent shift relative to the Fe{\sc\,ii} lines,
quite sufficient to simulate the variation of $\alpha$
(this shift can be easily estimated from recent
laboratory measurements \cite{Pickering}).

The method based on the fine splitting
of a line of the same ion species
is not affected by these two uncertainty sources.
We have studied
the fine splitting of the doublet lines of Si{\sc\,iv}, C{\sc\,iv}, 
Mg{\sc\,ii} and other ions, 
observed in spectra of distant quasars.
According to quantum electrodynamics, the relative splitting
of these lines $\delta\lambda/\lambda$ is proportional to 
$\alpha^2$ (neglecting small relativistic corrections, recently 
estimated by Dzuba et al.\ \cite{Dzuba}). We have selected 
the results of high-resolution observations \cite{Petitjean,VPI,Outram},
most suitable for an analysis of the variation of $\alpha$.
According to our analysis, presented elsewhere \cite{X99},
 the most reliable  estimate of the possible 
deviation of the fine-structure constant at $z = 2$--4 
from its present ($z = 0$) value:
\begin{equation}
\Delta\alpha/\alpha = 
      (-4.6\pm4.3\,[{\rm stat}]\pm1.4\,[{\rm syst}])\times10^{-5}.
\end{equation}
Thus, only an upper bound can be derived at present for 
the long-term variability of $\alpha$:
\begin{equation}
|\dot{\alpha}/\alpha| < 1.4\times10^{-14}{\rm~yr}^{-1}
\label{alpha-bound}
\end{equation}
(at the 95\% confidence level).

\section{Non-variability of $\mu$}
\label{sect-mu}

The dimensionless Born--Oppenheimer
constant $\mu = m_e/m_p$ approximately equals the ratio of the constant
of electromagnetic interaction $\alpha=e^2/\hbar c \approx 1/137$
to the constant of strong interaction $\alpha_s=g^2/\hbar c \sim 14$, 
where $g$ is the effective coupling constant calculated from the 
amplitude of $\pi$-meson--nucleon scattering at low energy.

An early limit on the possible variation of this constant, 
$|\dot\mu/\mu|<1.2\times10^{-10}{\rm~yr}^{-1}$, has been
derived from the concordance  of K--Ar and Rb--Sr
geochemical ages \cite{Yahil}. The first astrophysical bound \cite{Pagel},
based on the agreement between redshifts of atomic hydrogen and 
other lines in quasar absorption spectra,
turned out to be twice stronger.

Orders-of-magnitude more precise analysis
has become possible due to discovery \cite{LV85}
of a system of H$_2$ absorption lines
in the spectrum of quasar PKS 0528$-$250
at $z=2.811$. 
A study of this system yields information about
physical conditions and, in particular, 
the value of $\mu$ at this redshift (corresponding to 
the epoch when the Universe was several times younger than now).
A possibility 
of distinguishing between 
the cosmological redshift of spectral wavelengths 
and shifts due to a variation of $\mu$ arises from the 
fact that the electronic, vibrational, and rotational energies of 
H$_2$ each undergo a different dependence on the reduced mass of 
the molecule.  Hence 
comparing ratios of wavelengths $\lambda_i$ of various H$_2$
electron-vibration-rotational lines in a quasar spectrum at some 
redshift $z$ and in laboratory (at $z=0$), we can trace
variation of $\mu$. 
We have calculated \cite{SpSciRev,apj} sensitivity coefficients $K_i$
of the wavelengths $\lambda_i$ with respect
to possible variation of $\mu$ and applied a linear regression analysis to 
the measured redshifts of individual lines $z_i$ as function of $K_i$.
If the proton mass in the epoch of line formation
were different from the present value, the
measured $z_i$ and $K_i$ values would correlate:
\begin{equation}
{z_i\over z_k} = 
{(\lambda_i/\lambda_k)_z \over (\lambda_i/\lambda_k)_0} 
\simeq 
1+(K_i-K_k)\left({\Delta\mu\over\mu}\right).
\label{1}
\end{equation}
We have performed a $z$-to-$K$ regression analysis using
a modern high-resolution spectrum
of PKS 0528$-$250.
Eighty-two of the H$_2$ lines have been identified.
The resulting parameter estimate and $1\sigma$ uncertainty is
\begin{equation}
\Delta \mu/\mu = (-11.5\pm7.6{\rm\,[stat]}\pm1.9{\rm\,[syst]})
      \times 10^{-5}. 
\label{dmu}
\end{equation}
The $2 \sigma$ confidence bound on $\Delta \mu / \mu$ reads
\begin{equation}
 |\Delta \mu / \mu| < 2.0\times 10^{-4}. 
\label{limits}
\end{equation}
Assuming that the age of the Universe is $\sim 1.5\times10^{10}$ yr 
the redshift of the H$_2$ absorption system
$z=2.81080$ corresponds to the elapsed time $\approx 1.3\times10^{10}$ yr 
(in the standard cosmological model). 
Therefore we arrive at the restriction 
\begin{equation}
|\dot{\mu}/\mu|< 1.5\times 10^{-14}{\rm ~yr}^{-1}
\label{constraint}
\end{equation}
on the variation rate of $\mu$, averaged over 90\% of 
the lifetime of the Universe.

\section{Conclusions}
\label{sect-concl}

Despite the theoretical prediction that fundamental constants
of Nature should vary, 
no statistically significant variation of any of the constants
has been reliably detected up to date,
according to our point of view substantiated above.
The upper limits obtained indicate that the constants
of electroweak and strong interactions
did not significantly change over the last 90\% of the history
of the Universe.
The striking tightness of these limits is really astonishing
and has already ruled out some theoretical models
(see Refs.~\cite{SpSciRev,IPV,Pagel}).
A more elaborated theory (e.g., Ref.~\cite{DP}) 
cannot be ruled out yet, but its parameters
can be severely restricted (e.g., see Ref.~\cite{IPV}).
This shows that more precise measurements and observations
and their accurate statistical analyses are required
in order to detect the expected variations of the 
fundamental constants.

\vspace{\parsep}\noindent{\bf Acknowledgments.}
This work was performed in frames
of the Russian State Program ``Fundamental Metrology''
and supported by the grant RFBR 99-02-18232.

\newcommand{\ApJ}[1]{Astrophys.~J. {\bf #1}}
\newcommand{\AandA}[1]{Astron.\ Astrophys. {\bf #1}}
\newcommand{\JPB}[1]{J.\ Phys.\ B {\bf #1}}
\newcommand{\PRL}[1]{Phys.\ Rev.\ Lett. {\bf #1}}
\newcommand{\PR}[2]{Phys.\ Rev.\ #1 {\bf #2}}
\newcommand{\MNRAS}[1]{Mon.\ Not.\ Roy.\ Astron.\ Soc. {\bf #1}}

\end{document}